\documentclass[11pt,a4paper]{article}
\usepackage[utf8]{inputenc}
\usepackage{amsmath}
\usepackage{amssymb}
\usepackage{graphicx}
\usepackage{esint}
\usepackage[unicode=true,
 bookmarks=false,
 breaklinks=false,pdfborder={0 0 1},backref=section,colorlinks=false]
 {hyperref}

\makeatletter

\providecommand{\tabularnewline}{\\}

\pdfoutput=1

\usepackage{amsfonts}
\usepackage[font={small}]{caption}
\usepackage{epstopdf}%only when using miktex
\usepackage{color}
\usepackage[normalem]{ulem}

\usepackage{tabularx}
\usepackage{hhline}
\usepackage{booktabs}
\newcolumntype{Y}{>{\centering\arraybackslash}X}

\def\beq{\begin{equation}}
\def\eeq{\end{equation}}

\def\D{\mathrm d}

\def\-{\hphantom{-}}

\newcommand{\lowa}[  1]{#1 _{\mathrm a}}
\newcommand{\lowh}[  1]{#1 _{\mathrm h}}

\usepackage{geometry}
\title{Notes on ten-dimensional localized black holes and deconfined states in two-dimensional SYM}
\author{Martin Ammon, Michael Kalisch, Sebastian Moeckel  \\
\textit{\small{Theoretisch-Physikalisches Institut, Friedrich-Schiller-Universit\" at Jena,}} \\ 
\textit{\small{Max-Wien-Platz 1, D-07743 Jena, Germany}}}
\date{\today}

\makeatother

\begin{document}
\maketitle
\begin{abstract}
We numerically construct static localized black holes in ten spacetime
dimensions with one compact periodic dimension. In particular, we investigate
the critical regime in which the poles of the localized black hole
are about to merge. When approaching the critical region, the behavior of physical quantities is described by a single real valued exponent giving rise to a logarithmic
scaling of the thermodynamic quantities, in agreement with the theoretical
prediction derived from the double-cone metric. As a peculiarity,
the localized black hole solution in ten dimensions can be related
to the spatially deconfined phase of two dimensional $\mathcal{N}=\left(8,8\right)$
super Yang-Mills theory~(SYM) on a spatial circle. We use the localized black hole solutions to determine the SYM phase diagram. In particular, we compute the location of the first order
phase confinement/deconfinement transition and the related latent heat to unprecedented
accuracy.
\end{abstract}
%%%%%%%%%%%%%%%%% 
\newpage{}\tableofcontents{}

\section{\label{sec:Introduction}Introduction and Summary}
Throughout the last decades, the topic of black holes in higher dimensional spacetime with $D>4$
attracted a lot of attention. Special focus was devoted to the study
of $D$-dimensional black objects in spacetimes with one periodic
dimension, called Kaluza-Klein black holes. In this context there are various solutions to Einstein's
vacuum equations with different horizon topologies, such as black strings and localized black holes. The former are extended along the entire compact dimension while the latter are localized there.

Uniform black strings are unstable below a certain mass~\cite{Gregory:1993vy,Gregory:1994bj}. From this instability the branch of non-uniform black strings arises, which a number of authors have constructed in different dimensions
using perturbative and numerical techniques~\cite{Gubser:2001ac,Wiseman:2002zc,Sorkin:2004qq,Kleihaus:2006ee,Sorkin:2006wp,Headrick:2009pv,Figueras:2012xj,Kalisch:2015via,Kalisch:2016fkm,Dias:2017uyv}. However, thermodynamic arguments show that localized black holes are a more stable configuration for small masses. Such solutions have been constructed
perturbatively and numerically only in $D=5,\,6$~\cite{Myers:1986rx,Harmark:2003yz,Gorbonos:2004uc,Gorbonos:2005px,Wiseman:2002ti,Sorkin:2003ka,Kudoh:2003ki,Kudoh:2004hs,Headrick:2009pv,Kalisch:2017bin}
and recently in $D=10$~\cite{Dias:2017uyv}. Altogether this leads to an interesting and rather involved phase diagram of static Kaluza-Klein black holes. Good reviews summarizing the scientific progress in this realm can
be found in references~\cite{Kol:2004ww,Harmark:2005pp,Horowitz:2011cq,Kalisch:2018efd}.

In particular, recent work shed new light on this topic, showing that
the solution branches of non-uniform black strings and localized black
holes converge towards each other and that this transition is controlled
by complex critical exponents~\cite{Kalisch:2015via,Kalisch:2016fkm,Kalisch:2017bin}.
Interestingly, this critical behavior can be deduced from the so-called
double-cone metric, which Kol proposed as a local model of the transit
solution between non-uniform black strings and localized black holes~\cite{Kol:2002xz,Kol:2005vy}.

While these recent results only concern the dimensions $D=5,\,6$, the
present work concentrates on the construction of localized black holes in $D=10$ based on the methods presented in~\cite{Kalisch:2017bin}. Special attention
is devoted to the critical regime, where the poles of the horizons
are about to merge along the compact dimension. In particular, we show
that the critical regime in $D=10$ is approached in a qualitatively
different manner than for $D=5,\,6$, i.e.\ without oscillations.
Moreover, the obtained value of the real critical exponent is in excellent
agreement with the value that was predicted by Kol~\cite{Kol:2002xz,Kol:2005vy}.

In a second strand of the paper we investigate the phase diagram of maximally supersymmetric, two-dimensional super Yang-Mills theory~(SYM) on a spatial circle~$\mathbb{S}^{1}$ at strong coupling by the virtue of the AdS/CFT correspondence~\cite{Itzhaki:1998dd,Ammon:2015wua}. Due to the relation of the $D=10$ dimensional black hole and black string
solutions to thermal states of the SYM~\cite{Aharony:2004ig,Harmark:2004ws,Dias:2017uyv},
we are able to determine the phase diagram of the dual quantum field
theory. While black strings correspond to spatially confined phases, the localized black hole solution is dual to a deconfined phase. We locate the first order phase transition
between deconfined and confined phases in the microcanonical and canonical
ensemble with extraordinary accuracy. In addition, we calculate the
latent the heat of the phase transition and find a critical behavior
where the two meta-stable branches merge. All together, these quantities
provide valuable predictions for calculations within lattice quantum
field theory (see e.g.~\cite{Catterall:2010fx,Catterall:2017lub,Jha:2017zad}).

The paper is structured as follows: We review the physical setup for
localized black holes in ten-dimensional asymptotically flat spacetimes
in section~\ref{sec:LBHs_10D}. As in reference~\cite{Kalisch:2017bin},
the heart of our numerical implementation is a multi-domain pseudo-spectral
method. In section~\ref{subsec:Thermodynamics_and_critical_behavior}
we present our main results and compare the extracted critical exponent
with the theoretical predictions. As a second strand of the paper we study two-dimensional
maximally supersymmetric Yang-Mills theory using the conjectured gauge/gravity
duality in section~\ref{sec:SYM}. In particular, we determine its phase diagram in \ref{subsec:Thermodynamic-quantities}, and investigate the critical regime in the dual SYM in section~\ref{subsec:lessons}.

We provide supplementary material in appendix~\ref{sec:NumericalDetailsAndConvergence}
concerning the numerical implementation and the calculation of the phase transition points. Moreover, in appendix~\ref{sec:-SYM-on} we review the supergravity description of the two-dimensional maximally supersymmetric Yang-Mills theory and its regime of validity.

\textbf{Note added:} While this paper was being completed we became aware of upcoming work discussing similar issues \cite{Pau}.

\section{\label{sec:LBHs_10D}Localized black holes in ten dimensions}

This section is devoted to localized black hole solutions arising
from pure general relativity in ten dimensions with one compact spatial
dimension. The ansatz for their numerical construction is outlined
in subsection~\ref{subsec:Metric}, while we postpone a more detailed
description of the numerics to appendix~\ref{sec:NumericalDetailsAndConvergence}.
We utilized the numerical scheme developed in reference~\cite{Kalisch:2017bin}
and only made some minor adaptions. We present the results
of our computations in the localized black hole context in subsection~\ref{subsec:Thermodynamics_and_critical_behavior}.

\subsection{\label{subsec:Metric}Ansatz for the metric}

We consider ten-dimensional solutions of Einstein's vacuum field equations
$R_{\mu\nu}=0$, where one of the spatial dimensions is curled up
to a circle of size $L$. The simplest solution reads 
\beq \D s^2
= - \D t^2 + \D x^2 + x^2 \, \D \Omega^2_{7} + \D y^2 \, . \label{eq:background_metric}
\eeq 
Obviously, there is a spherical symmetry on the spatially extended
dimensions with the radial coordinate $x\in[0,\infty]$.
The additional coordinate $y$ is compact, $y\in[0,L]$, and periodically identified, i.e.
$y \simeq y+L$. This metric is a direct product of nine-dimensional Minkowski
spacetime and a circle~$\mathbb{S}^{1}$. Therefore, it serves as the background
metric, which all other solutions shall approach in the asymptotic
limit $x\to\infty$.

Keeping the spherical symmetry and, moreover, restricting ourselves
to static solutions, a general metric ansatz which incorporates the
required symmetries reads 
\beq \D s^2 = - \lowa{T} \, \D t^2 +
\lowa{A} \, \D x^2 + \lowa{B} \, \D y^2 + 2 \, \lowa{F} \,
\D x \, \D y + x^2 \lowa{S} \, \D \Omega^2_{7} \, . \label{eq:asymptotic_chart}
\eeq 
The five metric functions $\lowa T$, $\lowa A$, $\lowa B$,
$\lowa F$ and $\lowa S$ depend on $x$ and $y$. We obtain two asymptotic
charges, the mass $M$ and the relative tension $n$, from the subleading
behavior of the metric~\eqref{eq:asymptotic_chart} at infinity. In fact, these
two quantities are related to the coefficients $c_{t}$ and $c_{y}$
defined by 
\beq \lowa{T} \simeq 1 - c_t \, \frac{L^6}{x^{6}}
\, , \quad{}\lowa{B} \simeq 1 + c_y \, \frac{L^6}{x^{6}}
\, , \label{eq:asymptotic_corrections} \eeq 
in the following way
\begin{equation}
M=\frac{L^{7}\,\Omega_{7}}{16\pi G_{10}}\,\left(7c_{t}-c_{y}\right)\,,\quad n=\frac{c_{t}-7\,c_{y}}{7\,c_{t}-c_{y}}\,.\label{eq:mass}
\end{equation}
Note that the relative tension describes the force by which an object
tries to compress the compact dimension.

There are at least two types of static black hole solutions in Kaluza-Klein
theory: black strings, which are extended all over the compact dimension
and localized black holes, which are smaller than the compact dimension.
For the latter ones, being the subject of this work, the coordinates
used in the metric~\eqref{eq:asymptotic_chart} are not appropriate
to describe their near horizon behavior, since the horizon is some
curved contour in the $(x,y)$-plane. For this reason, we introduce
polar coordinates $(\varrho,\varphi)$ via $x=\varrho\sin\varphi$
and $y=\varrho\cos\varphi$, and rewrite the metric as 
\beq \D s^2
= -\kappa^2 (\varrho - \varrho_0)^2 \lowh{T} \, \D t^2 + \lowh{A}
\, \D \varrho^2 + \varrho^2 \lowh{B} \, \D \varphi^2 + 2\,
\varrho \lowh{F} \, \D \varrho \, \D \varphi + \varrho^2 \sin^2 \varphi \lowh{S} \, \D \Omega^2_7 \, . \label{eq:horizon_chart}
\eeq 
The functions $\lowh T$, $\lowh A$, $\lowh B$, $\lowh F$
and $\lowh S$ depend on $\varrho$ and $\varphi$, and are connected
with their counterparts of the metric~\eqref{eq:asymptotic_chart}
in a linear way. By the extraction of the term $\kappa^{2}(\varrho-\varrho_{0})^{2}$
from the $tt$-component of the metric we ensure that the horizon is
located at $x^{2}+y^{2}=\varrho^{2}=\varrho_{0}^{2}$. In fact, this
is only a gauge choice in a sense that we want the horizon to have
a spherical shape in the $(x,y)$ chart but the 'true' horizon shape
will not necessarily be exactly spherical. Thus, the parameter $\varrho_{0}$
does not have a specific physical meaning. In contrast, the surface
gravity $\kappa$ determines physically inequivalent solutions.

Our numerical approach to find localized black hole solutions relies
both on the metric~\eqref{eq:asymptotic_chart} and the metric~\eqref{eq:horizon_chart},
which we call the asymptotic and the near horizon chart, respectively.
This has the advantage of having coordinates and metric functions
that are well suited to different regions of the spacetime. We postpone
an outline of the numerical scheme to appendix~\ref{sec:NumericalDetailsAndConvergence}.

Besides the asymptotic charges there are some more physical quantities
of interest, in particular the temperature $T$ and the entropy $S$
of the black hole. While the temperature is simply related to the
surface gravity via $T=\kappa/(2\pi)$, the entropy is proportional
to the surface area of the horizon. In what follows we consider the
following dimensionless normalization of the physical quantities:
\begin{equation}
\tilde{M}=\frac{G_{10}\,M}{L^{7}}\,,\qquad\tilde{T}=TL\,,\qquad\tilde{S}=\frac{G_{10}\,S}{L^{8}}\,.
\end{equation}

\subsection{\label{subsec:Thermodynamics_and_critical_behavior}Thermodynamics
and critical behavior}

\begin{figure}[!ht]
\centering \includegraphics{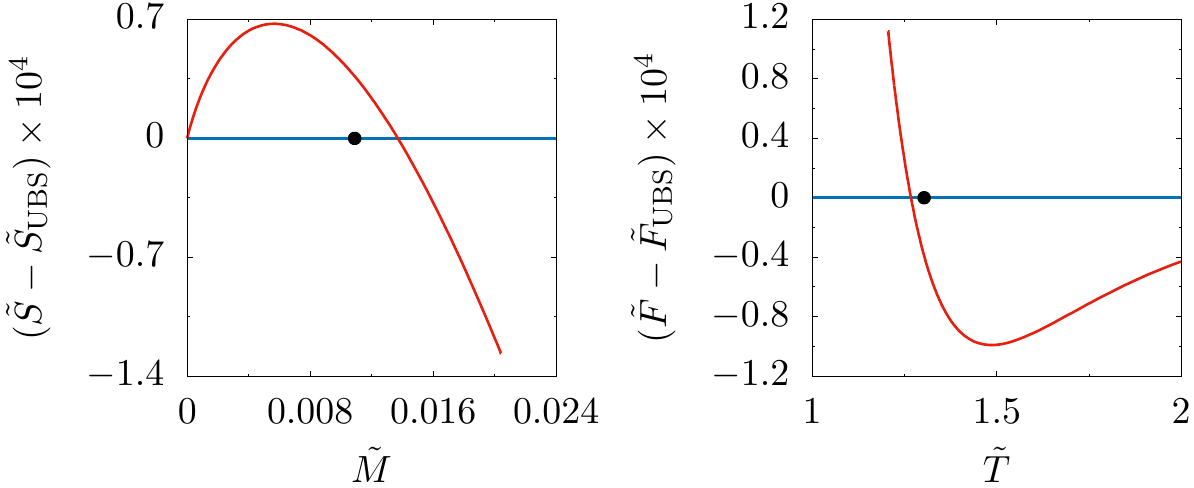} 
\caption{Phase diagram of 10-dimensional localized black holes in the microcanonical
(left) and canonical (right) ensemble. We plot the difference of entropy
and free energy, respectively, to the corresponding values of the
uniform black string, which is thus represented as the blue zero line
in these diagrams with the black circle indicating the solution where
the Gregory-Laflamme instability arises. The branch of localized black
holes, represented by the red line, is thermodynamically favored over
the uniform black strings for small masses or high temperatures.}
\label{fig:KKPhasediagrams} 
\end{figure}

We show the phase diagram of localized black holes in ten dimensions
in figure~\ref{fig:KKPhasediagrams} for the microcanonical and canonical
ensemble. The comparison with uniform black strings reveals that localized
black holes are thermodynamically favored for small masses and large
temperatures. Note that reference~\cite{Dias:2017uyv} did already
show this picture qualitatively and, moreover, included the non-uniform
black string results into the diagram.\footnote{In ten dimensions the non-uniform black string branch is at no point
thermodynamically favored. This changes in higher dimensions, see
references~\cite{Sorkin:2004qq,Figueras:2012xj,Emparan:2018bmi}.} However, we were able to extend the localized black hole solutions
much closer to the end point of this branch, where a transition to
non-uniform black strings is expected. Moreover, we are able to extract
the position of the first order phase transition (the point where
the uniform black string branch and the localized black hole branch
cross each other) from our data with high accuracy by distributing
the data points around this intersection on a Lobatto grid, see appendix
section~\ref{subsec:ObtainingThePhaseTransition} for more details.
In the microcanonical ensemble we obtain 
\beq \tilde M_\text{PT}
= 0.01369926356406(1) \, , \label{eq:MPTmicro} \eeq 
whereas in the canonical ensemble we get 
\beq \tilde T_\text{PT} = 1.267669090870(1)
\, . \label{eq:TPTmicro} \eeq 

\begin{figure}
\centering \includegraphics{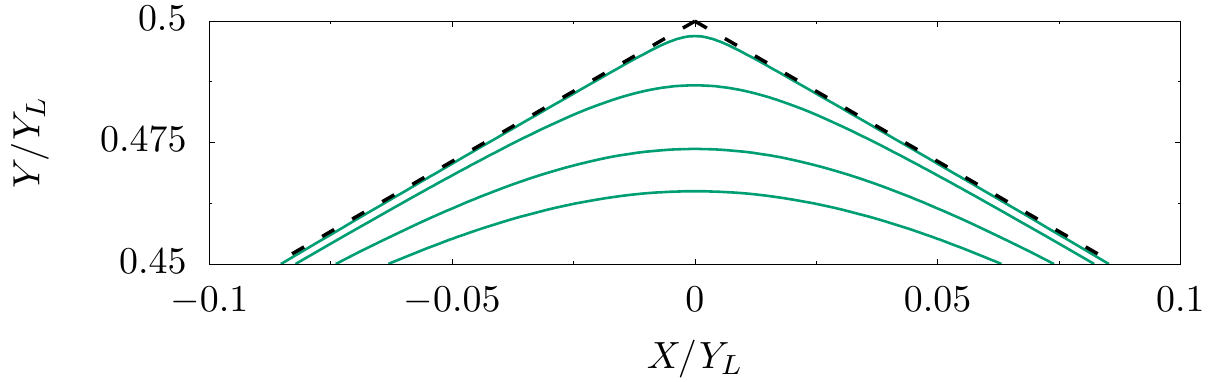} 
\caption{Local convergence towards the double-cone horizon shape. We embed the
horizons of different localized black holes solutions into nine-dimensional
flat space and display two-dimensional cross-sections disregarding
the seven-dimensional spherical symmetry. The non-trivial directions
$X$ and $Y$ correspond to our spacetime coordinates $x$ and $y$.
In the region close to the edge of the compact dimension, $Y\lesssim Y_{L}/2$,
we see that the shape predicted by the double-cone metric (dashed
line) is nicely approached by localized black hole horizons (green
lines) for increasing mass or decreasing temperature, respectively.}
\label{fig:approach_double_cone} 
\end{figure}

We now turn our attention to the end point of the localized black
hole branch. Following this branch it turns out that the black hole
horizon spreads more and more along the compact dimension. The transition
to non-uniform black strings is believed to be controlled by the so-called
double-cone metric~\cite{Kol:2002xz}, which is a local model of
the spacetime at the point where the poles of the localized black
hole touch each other. Indeed, figure~\ref{fig:approach_double_cone}
shows how the localized black hole horizon approaches the double-cone
locally. These horizon shapes were obtained by embedding the horizons
of different localized black hole solutions into flat space, cf.\ references~\cite{Kalisch:2017bin,Kalisch:2018efd}
for more details. 

Another interesting conjecture in the context of the double-cone metric
is the occurrence of critical exponents that control the thermodynamics
when the black hole/black string transition is approached~\cite{Kol:2002xz,Kol:2005vy}.
More concretely, we expect a certain physical quantity $p$ in this
regime to scale as 
\beq f - f_\text{c} = A \, Q^{-s_+} + B \,
Q^{-s_-} \, , \quad{}\text{with} \quad{}s_\pm = -\frac{D-2}{2}\left(
-1 \pm \text{i} \sqrt{\frac{8}{D-2} -1 } \right) \,
, \label{eq:criticalexponents} \eeq 
where $Q$ is a typical length
scale controlling the transition, $a$ and $b$ are constants and
$f_{\text{c}}$ is the value of the quantity $f$ at the transition.\footnote{Note that this result can be obtained by perturbing the double-cone metric and solving the corresponding ordinary
differential equations~\cite{Kol:2002xz}.} While for dimensions $D<10$ the exponents $s_{\pm}$ are complex
and hence lead to damped oscillations of physical quantities, 
they become purely real for $D\geq10$. For $D=5,6$ this expectation was explicitly
confirmed in reference~\cite{Kalisch:2017bin}. In the case considered
here, $D=10$, the exponents $s_{\pm}$ degenerate, $s_- = s_+$, and hence the scaling law \eqref{eq:criticalexponents} has to be modified as follows
\beq
f(Q) = f_\text{c} - Q^b ( a_1 + a_2 \, \log Q ) \, , \label{eq:FitAnsatz}
\eeq 
with $b=4$. Moreover, $a_{1}$ and $a_{2}$ are constants.\footnote{Note that the term $\log Q$ is a consequence of the degeneracy of the solution of the corresponding ordinary differential equation.}

\begin{table}
\noindent \begin{centering}
\caption{\label{tab:critical_exponents}Fit parameters for the function $f(Q)=f_{\text{c}}-Q^{b}\,(a_{1}+a_{2}\,\log Q)$
where $f$ stands for different physical quantities and $Q=L_{\mathcal{A}}/L$.
The fit parameters were determined using all data points with $Q\lesssim0.02$.}
\begin{tabular}{ccccc}
\hline 
$f$ & $f_{c}$ & $b$ & $a_{1}$ & $a_{2}$\tabularnewline
\hline 
\hline 
$\tilde{M}$ & 0.020404622 & 4.0009 & 0.2125 & -1.1380\tabularnewline
$n$ & 0.019873805 & 3.9999 & -7.8223 & -6.1737\tabularnewline
$\tilde{T}$ & 1.205852954 & 4.0006 & -0.5245 & 10.8728\tabularnewline
$\tilde{S}$ & 0.014764118 & 4.0009 & 0.1757 & -0.9440\tabularnewline
\hline 
\end{tabular}

\par\end{centering}
\end{table}

\begin{figure}
\centering \includegraphics{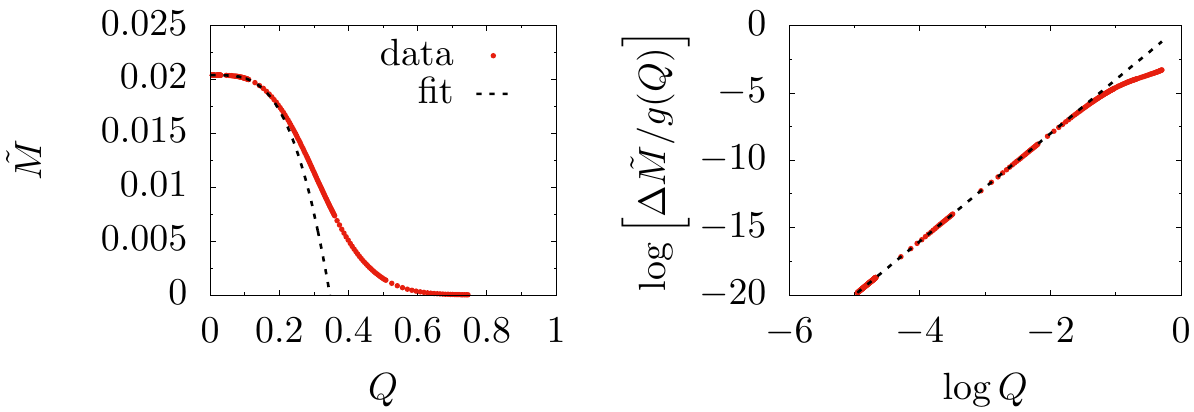} \caption{Scaling of mass. We show the rescaled mass $\tilde{M}$ as a function
of the normalized proper distance between the poles $Q=L_{\mathcal{A}}/L$
(left). To illustrate the remarkable agreement of data points and
fit, we plot the mass difference $\Delta\tilde{M}=\tilde{M}_{0}-\tilde{M}$
divided by $g(Q)=a_{1}+a_{2}\,\log Q$
against $Q$ in a double logarithmic diagram (right). }
\label{fig:critical_exponents} 
\end{figure}

In order to check this conjecture we follow reference~\cite{Kalisch:2017bin}
and fit our data for the ten-dimensional localized black holes with the ansatz~\eqref{eq:FitAnsatz} leaving
$f_{\text{c}}$, $b$, $a_{1}$ and $a_{2}$ as free parameters to
be determined by the fit routine. We use the proper distance between
the poles $L_{\mathcal{A}}$ as the length scale to describe the transition,
as $L_{\mathcal{A}}=0$ when the transition is reached. Accordingly,
we set $Q=L_{\mathcal{A}}/L$. Table~\ref{tab:critical_exponents}
shows the obtained fit parameter values for different physical quantities.
Most importantly, the theoretical predicted value $b=4$ is confirmed
to great accuracy. Moreover, we display the good agreement of data
points and fit exemplarily for the mass in figure~\ref{fig:critical_exponents}.

\section{\label{sec:SYM}Thermal states of $\mathcal{N}=(8,8)$ SYM and localized black holes}

Via the renowned gauge/gravity duality conjecture, Kaluza-Klein black holes in $D=10$ are related to thermal
states of a two-dimensional $\mathcal{N}=\left(8,8\right)$ supersymmetric
Yang-Mills (SYM) theory compactified to a circle~$\mathbb{S}^{1}$
with gauge group~$SU(N)$ in the large $N$ limit. In particular, localized black
holes correspond to a spatially deconfined phase within the SYM, while
 black strings are related to
spatially confined phases. 

Note that the aforementioned SYM can be
characterized by three dimensionless quantities: The rank of the gauge
group~$N$, the dimensionless 't~Hooft coupling constant\footnote{Note that the Yang-Mills coupling constant $g_{YM}$ has the dimension
of energy in two-dimensions.} $\lambda=Ng_{YM}^{2}L^{2}$ and the dimensionless temperature $\mathcal{T}$
given by the product of the ordinary temperature and the length of
the circle $L$. In the following, we denote the (dimensionless) thermodynamic
quantities of the SYM by $\mathcal{U}$ for the energy, $\mathcal{T}$
for the temperature and $\mathcal{S}$ for the entropy.

According to~\cite{Aharony:2004ig,Harmark:2004ws,Dias:2017uyv},
the localized black holes and black strings in $D=10$ dimensional asymptotically flat spacetime with one compact periodic dimension discussed in section \ref{sec:LBHs_10D} can be related to the gravitational dual solution of SYM states by employing the the following solution generating technique:
\begin{enumerate}
\item[(i)] First, we lift the $D=10$ dimensional solutions of vacuum Einstein equations discussed in section \ref{sec:LBHs_10D} to $D=11$ dimensions and perform a boost in the new coordinate, followed by a subsequent Kaluza-Klein reduction.

The result is a solution in type IIA supergravity. Concretely, the
localized black hole solutions in $D=10$ with $\mathbb{R}^{1,8}\times\mathbb{S}^{1}$
asymptotics correspond to localized D0-branes in type IIA supergravity.
\item[(ii)] As a next step a T-duality transformation is applied, which converts
the type IIA supergravity solution into a type IIB supergravity solution.
\item[(iii)] As a last step we take the decoupling limit between the string and
gravitational length scales on the type IIB gravity side, which corresponds
to taking the limit $N\rightarrow\infty$ with $\lambda$ fixed on
the SYM side, and to consider the large $\lambda$ limit in a second
step.
\end{enumerate}
The details of the decoupling limit are rather complicated and
we refer the reader to the references~\cite{Aharony:2004ig,Harmark:2004ws,Dias:2017uyv} and appendix ~\ref{sec:-SYM-on}
for a thorough description of the underlying limits and an analysis on the validity of the supergravity description of the SYM states. \par
We determine the phase diagram of the $\mathcal{N}=\left(8,8\right)$ SYM compactified to a circle~$\mathbb{S}^{1}$  in section~\ref{subsec:Thermodynamic-quantities} by applying the solution generating to the ten-dimensional localized black holes in asymptotically flat spacetime. In particular we locate the first order phase transition to very high accuracy. Moreover, in~\ref{subsec:lessons} we interpret the critical regime between localized black holes and non-uniform black strings as an emergent critical scaling behavior related to the first order phase transition  where the the two meta-stable branches merge.

\subsection{\label{subsec:Thermodynamic-quantities}Thermodynamic quantities}

As a result of the solution generating procedure, we can relate
the thermodynamic quantities of the SYM to the thermodynamic properties
of the Kaluza-Klein black holes. In particular, considering the normalized
quantities $\mathcal{\tilde{U}}=\mathcal{U}\cdot\lambda^{2}/N^{2}$,
$\mathcal{\tilde{T}}=\mathcal{T}\cdot\lambda^{1/2}$ and $\mathcal{\tilde{S}}=\mathcal{S}\cdot\lambda^{3/2}/N^{2}$,
we obtain (cf. reference \cite{Dias:2017uyv})
\begin{equation}
\mathcal{\tilde{U}}=64\,\pi^{4}\left(2\tilde{M}-\tilde{S}\tilde{T}\right)\,,\qquad\mathcal{\tilde{T}}=4\sqrt{2}\pi\tilde{S}^{1/2}\tilde{T}^{3/2}\,,\qquad\mathcal{\tilde{S}}=16\sqrt{2}\pi^{3}\sqrt{\frac{\tilde{S}}{\tilde{T}}}\,.
\end{equation}
The free energy $\mathcal{F}$ of the canonical ensemble and its normalized
version $\mathcal{\tilde{F}}=\mathcal{F}\cdot\lambda^{2}/N^{2}$ are
given by $\mathcal{F}=\mathcal{U}-\mathcal{S}\mathcal{T}$ and $\mathcal{\tilde{F}}=\mathcal{\tilde{U}}-\mathcal{\tilde{S}}\mathcal{\tilde{T}}$.

\begin{figure}
\centering \includegraphics{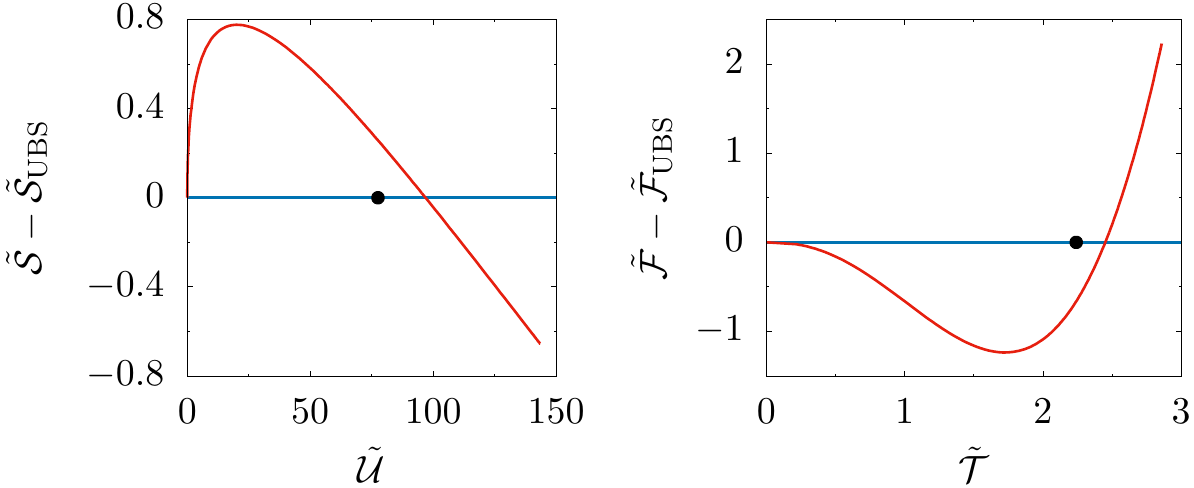} \caption{\label{fig:SYMPhasediagrams}Phase diagram of $\mathcal{N}=(8,8)$
SYM on a circle with length $L$ in the microcanonical (left) and
canonical (right) ensemble. We plot the difference of entropy and
free energy, respectively, to the corresponding values of the uniform
branch, which is thus represented as the blue zero line in these diagrams
with the black circle indicating the solution where the Gregory-Laflamme
instability arises. The localized branch, represented by the red line,
is thermodynamically favored over the uniform branch for small masses
or small temperatures.}
\end{figure}

Figure~\ref{fig:SYMPhasediagrams} shows the phase diagrams of the
microcanonical and canonical ensembles of the uniform and localized phases of the $\mathcal{N}=\left(8,8\right)$
SYM. For the microcanonical ensemble, we see that the localized phase
is thermodynamically preferred over the uniform phase up to some threshold
value of the normalized internal energy $\tilde{\mathcal{U}}_{PT}$,
where the uniform phase starts to dominate. There is a first order
phase transition, where the entropy of the uniform phase exceeds the
entropy of the localized branch. We have a similar picture when considering
the canonical phase diagram. Here, lower values of the free energy
$\mathcal{\tilde{\mathcal{F}}}$ correspond to the thermodynamically
preferred phase. Accordingly, we see that the localized phase is dominating
for small values of the normalized temperature $\tilde{\mathcal{T}}$.
As before, the uniform phase becomes thermodynamically preferred at
some threshold value $\tilde{\mathcal{T}}_{PT}$. We remark that including
the SYM phases corresponding to non-uniform black strings will not alter
this picture of thermodynamic stability, since the related branch
is thermodynamically inferior for all configurations, as can be seen
from the data presented in reference~\cite{Dias:2017uyv}.

With the procedure described in appendix section~\ref{subsec:ObtainingThePhaseTransition}
we determine the first order phase transition between the localized
and the uniform phase in the microcanonical ensemble to be at 
\beq
\mathcal{\tilde U}_\text{PT} = 96.9053906163(1) \, . \label{eq:UsymPTmicro}
\eeq
For the canonical ensemble we find 
\beq \mathcal{\tilde
T}_\text{PT} = 2.451118333749(1)  \label{eq:TsymPTcano} \, . \eeq
Both values are in good agreement with reference~\cite{Dias:2017uyv} which determines $\mathcal{\tilde U}_\text{PT} \approx 97.067$ and
$\mathcal{\tilde{T}}_{\text{PT}}\approx2.451$. Moreover, the latent
heat associated with the first order phase transition between localized black holes and uniform black strings (UBS) is given by

\begin{subequations} 
\begin{align}
\Delta\mathcal{\tilde{Q}} & =\mathcal{\tilde{T}}_{\text{PT}}\,\cdot\,\left(\mathcal{\tilde{S}}_{\mathrm{UBS}}-\mathcal{\tilde{S}}\right)\Big|_{\mathcal{\tilde{T}}=\mathcal{\tilde{T}}_{\text{PT}}}\label{eq:LatentHeata}\\
 & =\mathcal{\tilde{T}}_{\text{PT}}\,\cdot\,\frac{\partial(\mathcal{\tilde{F}}-\mathcal{\tilde{F}}_{\mathrm{UBS}})}{\partial\mathcal{\tilde{T}}}\Big|_{\mathcal{\tilde{T}}=\mathcal{\tilde{T}}_{\text{PT}}}\label{eq:LatentHeatb}
\end{align}
\end{subequations} where $\mathcal{\tilde{S}}(\mathcal{\tilde{T}})$
is the normalized entropy associated with the localized black holes.
Utilizing again the procedure described in appendix section~\ref{subsec:ObtainingThePhaseTransition}
we determine the latent heat to be \beq \Delta \mathcal{\tilde
Q} = 9.47738683316(1) \, . \label{eq:LatentHeatValue} \eeq Note
that this value was obtained from equation~\eqref{eq:LatentHeata}.
We also evaluated equation~\eqref{eq:LatentHeatb} and only found
deviations within the last two digits compared to the value given in equation~\eqref{eq:LatentHeatValue},
which is completely expected since in this case we have to perform
a numerical derivative of the free energy.

Again, when approaching the end point of the localized branch, the
physical quantities show a scaling behavior, reminiscent of the one for the localized black holes in ten dimensional Kaluza-Klein geometries. While this is expected
from the gravitational point of view, cf.\ reference~\cite{Kol:2005vy},
this is surprising from the field theory perspective. To our best knowledge, we are not aware of results in the literature concerning an emergent scaling behaviour with real critical exponents when the two meta-stable branches merge into each other. 

In table~\ref{tab:critical_exponents_SYM} we list these unexpected critical exponents (see third column) which were obtained by fitting our data with the same ansatz~\eqref{eq:FitAnsatz}
as we used for the localized black holes. 
\begin{table}
\noindent \centering{}\caption{\label{tab:critical_exponents_SYM}Fit parameters for the function
$f(Q)=f_{\text{c}}-Q^{b}\,(a_{1}+a_{2}\,\log Q)$ where $f$ stands
for different physical quantities and $Q=L_{\mathcal{A}}/L$. The
fit parameters were determined using all data points with $Q\lesssim0.02$.}
\begin{tabular}{ccccc}
\hline 
$f$ & $f_{c}$ & $b$ & $a_{1}$ & $a_{2}$\tabularnewline
\hline 
\hline 
$\mathcal{\tilde{U}}$ & 143.42290 & 4.0009 & 1362 & -8100\tabularnewline
$\mathcal{\tilde{T}}$ & 2.8593686 & 4.0008 & 15.46 & -52.60\tabularnewline
$\mathcal{\tilde{S}}$ & 77.632071 & 4.0008 & 480.1 & -2831\tabularnewline
$\mathcal{\tilde{F}}$ & -78.555803 & 4.0010 & -1188 & 4088\tabularnewline
\hline 
\end{tabular}
\end{table}

\subsection{Lessons for the dual SYM}\label{subsec:lessons}

We obtained the phase diagrams of the localized and uniform phases
of the two-dimensional $\mathcal{N}=\left(8,8\right)$ supersymmetric
SYM theory compactified to a circle~$\mathbb{S}^{1}$ with gauge
group~$SU(N)$ in the large $N$ limit from the corresponding localized black hole
and uniform black string solutions. The localized branch was found to be predominating for
small energies or temperatures, whereas the uniform phase becomes
thermodynamically favored at some threshold values of the energy $\tilde{\mathcal{U}}_{PT}$
or temperature $\tilde{\mathcal{T}}_{PT}$. We obtained the threshold
values with unprecedented accuracy and additionally computed the latent
heat $\Delta\tilde{\mathcal{Q}}$ of the related first order phase
transition.

Especially the values $\tilde{\mathcal{U}}_{PT}$, $\tilde{\mathcal{T}}_{PT}$
and $\Delta\tilde{\mathcal{Q}}$ are the basis for a comparison of
our results with regarding data from quantum lattice calculations.
We refer the reader to the references \cite{Catterall:2010fx,Catterall:2017lub,Jha:2017zad}
for lattice results that show indications of a first-order phase transition.
We remark, that possible phase transitions within this theory, are
detected within lattice calculations by considering the Polyakov loop
$P_{L}$ for a closed curve $\mathcal{C}_{L}$ along the periodic
spatial direction 
\begin{align}
P_{L} & =\frac{1}{N}\left\langle \textrm{Tr}\,\mathcal{P}\exp\left(\text{i}\oint_{\mathcal{C}_{L}}A\right)\right\rangle \,,\label{eq:P_L}
\end{align}
where $\mathcal{P}$ indicates the usual ordering prescription for
Polyakov and Wilson loops. $P_{L}$ is an order parameter for a spatial
confinement/deconfinement phase transition: $P_{L}\neq0$ indicates
a deconfined spatial behavior, while $P_{L}=0$ signals a confined
spatial behavior. Within the dual supergravity theory, the localized black hole phase corresponds to a spatially deconfined phase
with non-zero Polyakov loop $P_{L}\neq0$ while the black string solutions
are related to a spatially confined phases with $P_{L}=0$.

A more refined observable is the eigenvalue distribution of the Polyakov loop $P_L$ \cite{Aharony:2004ig} on the complex unit circle. Note that the eigenvalue distribution is continuous in the large $N$ limit. This observable allows us to distinguish between states dual to localized black holes as well as non-uniform and uniform black strings. While uniform black strings are dual to a state with a homogeneous eigenvalue distribution on the complex unit circle, the non-uniform black strings correspond to a non-uniform eigenvalue distribution which is spread over the entire unit circle, i.e.\ for black strings we have the eigenvalues $\exp(\text{i}\varphi)$ for all $\varphi \in [-\pi, \pi]$.  

In contrast, for the state dual to the localized black hole, we only have eigenvalues for $\varphi \in [-\varphi_0, \varphi_0],$ where $\varphi_0 < \pi$. In other words, the eigenvalue distribution is only nonzero for $|\varphi|\leq \varphi _0$. Hence, the state corresponding to the limiting case $\varphi_0 \rightarrow \pi$ is dual to the merger solution in the gravitational theory, approached from the branch of the localized black holes. We expect that the scaling law \eqref{eq:FitAnsatz}, with $Q \sim \pi - \varphi_0$ and the critical exponents as reported in table \ref{tab:critical_exponents_SYM}, should emerge while interpolating between the localized and non-homogeneous eigenvalue distributions. 

\section*{Acknowledgments}
MK and SM acknowledge financial support by the Deutsche Forschungsgemeinschaft
(DFG) GRK 1523/2.

\appendix

\section{\label{sec:Numerical-details-and}Numerical details and convergence}

\label{sec:NumericalDetailsAndConvergence}

Reference~\cite{Kalisch:2017bin} gives a comprehensive discussion
of how to construct localized black hole solutions in five and six
spacetime dimensions very accurately. The setup that we utilized for the
ten-dimensional case heavily relies on this approach. In subsection~\ref{subsec:OverallScheme}
we only want to emphasize the corner stones of our numerical implementation
while referring to reference~\cite{Kalisch:2017bin} for more details.
However, to extract the asymptotic charges accurately, we had to introduce some new techniques, which we explain in subsection~\ref{subsec:ExtractionOfAsymptoticCharges}.
In subsection~\ref{subsec:TestOfNumericalResults} we discuss
the accuracy and the convergence of the numerical solutions. Finally, subsection~\ref{subsec:ObtainingThePhaseTransition} shows how we obtained the highly accurate values regarding the phase transition from our data. 

\subsection{\label{subsec:OverallScheme}Overall scheme}

We utilize the DeTurck method~\cite{Headrick:2009pv} in order to
find numerical solutions to Einstein's vacuum equations in the given
context, see references~\cite{Wiseman:2011by,Dias:2015nua} for reviews.
Hence, rather than solving these equations directly, we want to find
solutions to the Einstein-DeTurck equations 
\beq R_{\mu\nu} -
\nabla_{(\mu}\xi_{\nu )} = 0 \, , \label{eq:EinsteinDeTurck}
\eeq 
with the DeTurck vector field $\xi$ defined by 
\begin{equation}
\xi^{\mu}=g^{\alpha\beta}\left(\Gamma_{\alpha\beta}^{\mu}-\bar{\Gamma}_{\alpha\beta}^{\mu}\right)\,.\label{eq:DeTurckvector}
\end{equation}
While $\Gamma$ is the usual Christoffel connection obtained from
the desired spacetime metric $g$, $\bar{\Gamma}$ is the Christoffel
connection associated with an unphysical reference metric $\bar{g}$
that only needs to exhibit the same causal structure and boundary conditions
as $g$. If this is the case, a solution $g$ of the Einstein-DeTurck
equations also satisfies Einstein's vacuum equations, at least in
the static case considered here~\cite{Figueras:2011va}. Nevertheless,
for a numerical solution $g$ it is always a good idea to check if
the DeTurck vector is sufficiently close to zero, which ensures that
the additional term in the Einstein-DeTurck equations vanishes.

To construct an appropriate reference metric we follow the lines of
reference~\cite{Kalisch:2017bin}: Observing that the background
metric~\eqref{eq:background_metric} already satisfies the right boundary
conditions on all boundaries except the horizon, we simply take this
metric as reference but only for $x^{2}+y^{2}=\varrho^{2}\geq\varrho_{1}^{2}$,
see also reference~\cite{Headrick:2009pv}. Within $\varrho_{0}\leq\varrho\leq\varrho_{1}$
we construct the reference by matching it with the background metric
at $\varrho=\varrho_{1}$ and with a ten-dimensional Schwarzschild-Tangherlini
metric at $\varrho=\varrho_{0}$, see reference~\cite{Kalisch:2017bin}
for more details.

\begin{figure}[ht]
\centering \includegraphics{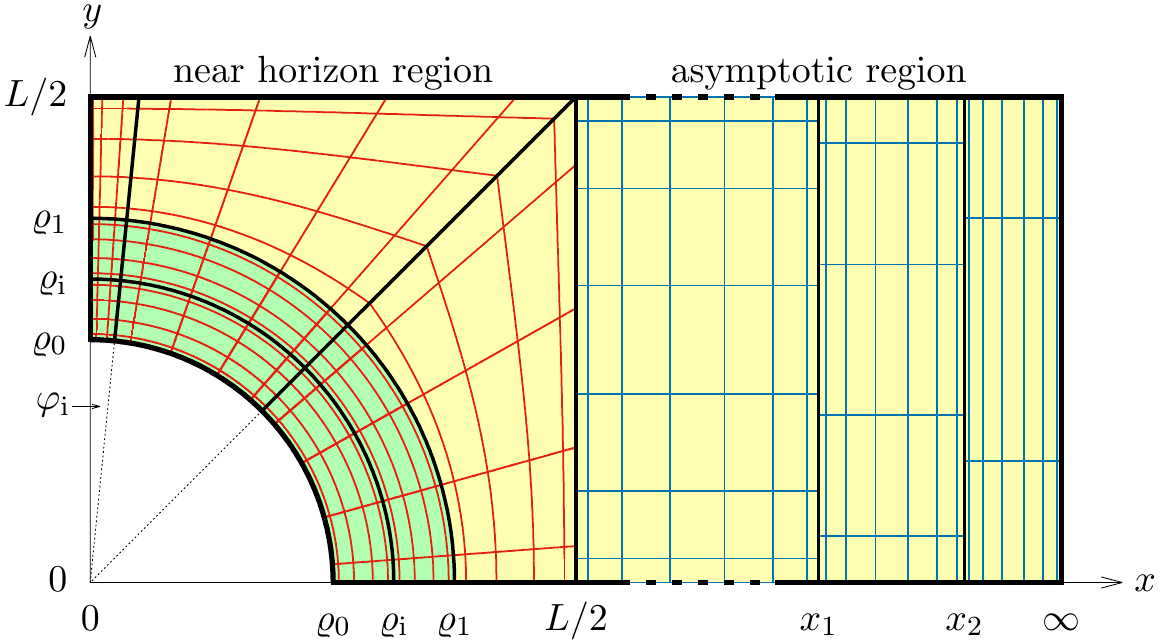} \caption{Grid structure for the construction of localized black hole solutions.
For $x\leq L/2$ the computations are carried out with respect to
the near horizon chart~\eqref{eq:horizon_chart} (indicated by red
grid lines) while for $x\geq L/2$ the asymptotic chart~\eqref{eq:asymptotic_chart}
is used (indicated by blue grid lines). For $\varrho\leq\varrho_{1}$
(green shaded region) we define a $\varphi$-independent reference
metric that only contains powers of $\varrho$ by matching with the
the ten-dimensional Schwarzschild-Tangherlini metric at the horizon
$\varrho=\varrho_{0}$ and with the Kaluza-Klein background metric
at $\varrho=\varrho_{1}$. Consequently, for $\varrho>\varrho_{1}$
(yellow shaded region) we simply use the Kaluza-Klein background as
a reference. We introduce additional inner boundaries along $\varrho=\varrho_{\text{i}}$,
$\varphi=\varphi_{\text{i}}$, $x=x_{1}$ and $x=x_{2}$ in order
to be able to increase the numerical resolution especially near the
horizon $\varrho=\varrho_{0}$, the exposed axis $\varphi=0$ and
infinity $x\to\infty$. For the majority of our calculations we used
the following parameter values: $L=8$, $\varrho_{0}=2$, $\varrho_{\text{i}}=2.5$,
$\varrho_{1}=3$, $\varphi=0.1$, $x_{1}=L$ and $x_{2}=5\,L$. The
reference metric itself can be used to construct an initial guess
for the Newton-Raphson scheme with the above parameter values and
$\kappa=1.4$. Once a first solution is found, different solutions
are obtained by slightly perturbing $\kappa$ and using the old solution
as an initial guess.}
\label{fig:Grid} 
\end{figure}

Our numerical approach relies on a pseudo-spectral method. In particular,
we approximate all functions with a truncated series of Chebyshev
polynomials of the first kind, while we demand that this approximation
is exact on Lobatto grid points. With this we are able to utilize
the Newton-Raphson method in order to solve the differential equations
on the grid. The rate of convergence of a Chebyshev series heavily
relies on smoothness properties of the underlying function. For this
reason we decompose the domain of integration into several subdomains
and perform appropriate coordinate transformations. The resulting
grid setup is discussed in figure~\ref{fig:Grid} but we refer to
reference~\cite{Kalisch:2017bin} for more details. 

The domain of integration consists of five outer boundaries: the asymptotic
boundary ($x\to\infty$), start and mid point of the compact dimension
($y=0$ and $y=L/2$), the symmetry axis ($x=0$) and the horizon
($x^{2}+y^{2}=\varrho_{0}^{2}$). We divide the domain of integration
into an asymptotic region ($x\geq L/2$) and a near horizon region
($x\leq L/2$). In the former we consider the metric functions of
the asymptotic chart~\eqref{eq:asymptotic_chart}, while in the latter
it is more convenient to work with the polar coordinates in the near
horizon chart~\eqref{eq:horizon_chart}.

The numerical method requires boundary conditions for all functions
in each subdomain. On inner boundaries we simply demand equality of
the metric function values and their normal derivatives with respect
to neighboring subdomains. Conditions on the five outer boundaries
mentioned above are obtained from regularity and symmetry requirements
of the metric and are derived from the field equations itself, see
reference~\cite{Kalisch:2017bin} for the explicit conditions or
reference~\cite{Dias:2015nua} for a more general discussion of boundary
conditions in the context of the DeTurck method. However, an exception
is the asymptotic boundary, where the metric shall approach the background
and therefore Dirichlet conditions are usually employed. As mentioned
before, since we want to extract the asymptotic charges quite accurately,
we use a more sophisticated approach here, which we explain in the
next subsection.

\subsection{\label{subsec:ExtractionOfAsymptoticCharges}Extraction of asymptotic
charges}

To obtain the asymptotic charges, mass and tension, we need to extract
the coefficients $c_{t}$ and $c_{y}$ from the functions $\lowa T$
and $\lowa B$, cf.\ equation~\eqref{eq:asymptotic_corrections}.
Of course, we cover the integration domain up to infinity with an
appropriate coordinate transformation that compactifies the asymptotic
boundary to a finite coordinate value. This coordinate transformation
reads $x(s)=L/(1-s)$ where $s\in[-1,1]$ covers the whole asymptotic
region $x\in[L/2,\infty]$. Consequently, in principle it is possible
to get the asymptotic coefficients $c_{t}$ and $c_{y}$ from the
sixth derivative of the functions $\lowa T$ and $\lowa B$ with respect
to $s$. However, since each numerical derivative is accompanied with
some small errors, we cannot expect the sixth derivative to be reasonable
accurate. 

This is in contrast to the five- or six-dimensional approach, where
only the first or second derivatives are involved and an accurate
determination of the asymptotic coefficients is possible, see reference~\cite{Kalisch:2017bin}.

Therefore, in the ten-dimensional case we employ the following approach:
We write the functions in the asymptotic chart $\lowa X=\{\lowa T,\lowa A,\lowa B,\lowa S\}$
and $\lowa F$ as 
\begin{equation}
\lowa X=1+\frac{(1-s)^{5}}{32}\tilde{\lowa X}\quad\text{and}\quad\lowa F=\frac{(1-s)^{5}}{32}\tilde{\lowa F}\,,\label{eq:asymptotic_extraction}
\end{equation}
and now solve for the new functions $\tilde{\lowa X}$ and $\tilde{\lowa F}$.
In this way we obtain the asymptotic coefficients $c_{t}$ and $c_{y}$
only from the first derivatives of the functions $\tilde{\lowa T}$
and $\tilde{\lowa B}$ at $s=1$. Note that the ansatz~\eqref{eq:asymptotic_extraction}
incorporates the leading asymptotic behavior, namely that the spacetime
approaches the Kaluza-Klein background at $s=1$ where $\lowa X=1$
and $\lowa F=0$. The boundary conditions on the newly defined functions
are $\tilde{\lowa X}=\tilde{\lowa F}=0$.

Let us make a few technical comments on the above described trick.
First, one could ask why we do not extract one more power of $(1-s)$
from the functions leading to a scenario where we can read off the
asymptotic coefficients directly from the values of the redefined
functions at $s=1$ without performing any derivative. The reason
is that this may lead to some more complicated conditions on the functions
at $s=1$ as it is the case in five and six dimensions, see reference~\cite{Kalisch:2016fkm}
where in the black string setup a rather involved decomposition of
the functions near infinity was utilized. With the approach described
here we circumvent these technical obstacles.

The second question one could ask is why do we extract the appropriate
powers of $(1-s)$ from all metric functions and not only from $\lowa T$
and $\lowa B$, since we are only interested in their sixth derivatives.
If we would do so, this would probably worsen the accuracy of the
extracted values of the asymptotic coefficients because near the asymptotic
boundary $s=1$ the functions $\tilde{\lowa T}$ and $\tilde{\lowa B}$
would be suppressed in the field equations by a factor of $(1-s)^{5}$.
As a result, conditions involving the asymptotic coefficients would
be suppressed considerably. This is a problem in numerical calculations
due to finite machine precision and we thus cannot expect the extracted
values of the asymptotic coefficients to be very accurate. However,
with the ansatz~\eqref{eq:asymptotic_extraction} we ensure that
we can extract an appropriate power of $(1-s)$ from each field equation
avoiding the suppression of the crucial terms.

Finally, we note that the factors of 32 appearing in the ansatz~\eqref{eq:asymptotic_extraction}
are chosen in order to normalize the functions and field equations
at $s=-1$.

\subsection{\label{subsec:TestOfNumericalResults}Test of numerical results}

As usual, in order to assess the reliability of the numerical solutions,
we need to study the convergence of the numerical solution scheme in dependence of the grid resolution.
In particular, we compare the values of the solution functions for
increasing resolution with a reference solution at high resolution, where we denote the difference as the residual.
Furthermore, for all solutions we display the deviation from Smarr's
relation, cf.\ references~\cite{Kol:2003if,Harmark:2003dg}, with
the relevant physical quantities obtained from the corresponding numerical
data. In theory, the resulting errors should decrease as the resolution
increases at least up to some remaining round-off error.

\begin{figure}[ht]
\centering \includegraphics{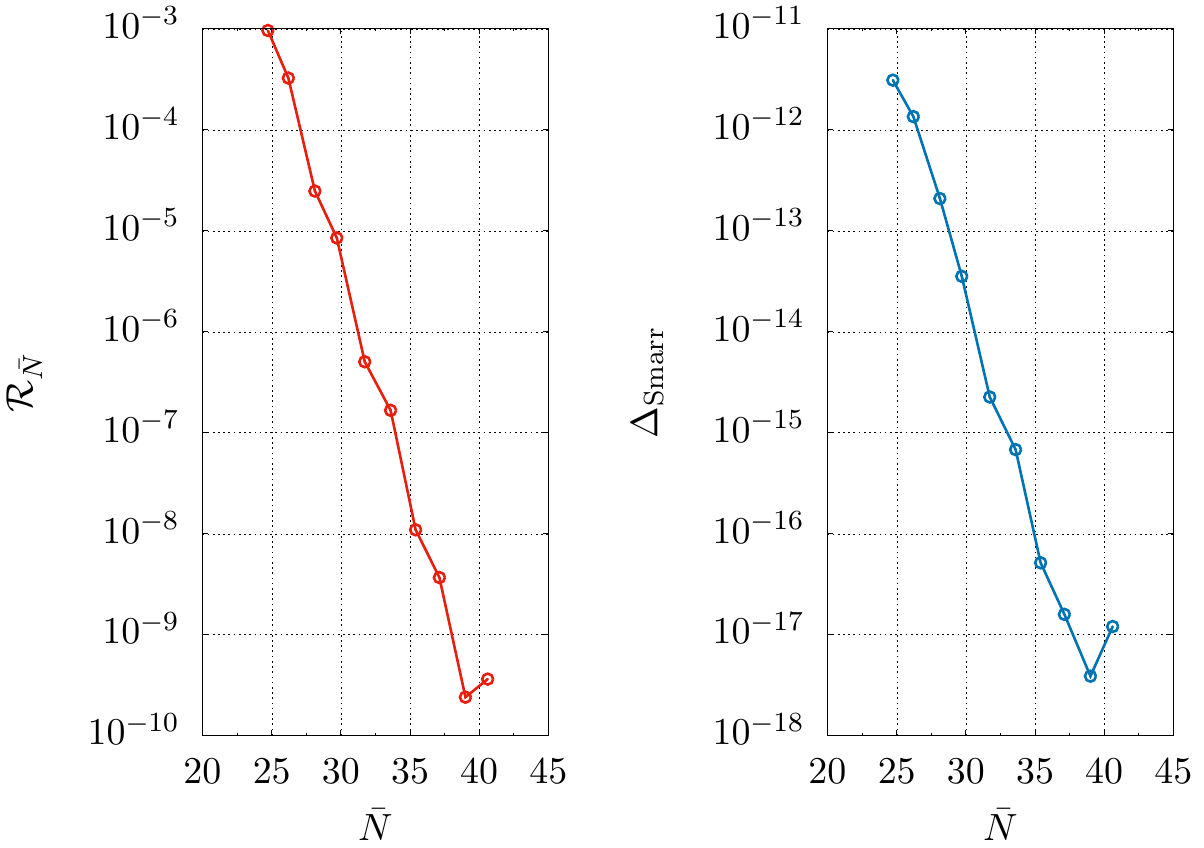} \caption{Convergence of the numerical solutions for a localized black hole
with inter polar distance $L_{\mathcal{A}}/L\approx0.00703$. For
different resolutions $\bar{N}$ we show the maximal difference $\mathcal{R}_{\bar{N}}$
to a reference solution with high resolution (left) and the deviation
from Smarr's relation $\Delta_{\text{Smarr}}=|(D-2)\tilde{T}\tilde{S}-(D-3-n)\tilde{M}|$ (right). The mean resolution $\bar{N}$ is averaged over all domains
and directions.}
\label{fig:Accuracy} 
\end{figure}

Exemplary, in figure~\ref{fig:Accuracy} we show the convergence
of the residual for the configuration with inter polar distance $L_{\mathcal{A}}/L\approx0.00703$,
which is the solution closest to the transition that we presented
here.\footnote{We note that we were able to construct solutions with smaller $L_{\mathcal{A}}/L$,
but the accuracy of these solutions dropped down considerably.} As expected, we see a nice convergence of the respective errors.

Finally, we note that the maximum of the non-trivial components of
the DeTurck vector $\xi$, cf.\ equation~\eqref{eq:DeTurckvector},
always remained below $10^{-10}$ for all solutions we constructed.

\subsection{\label{subsec:ObtainingThePhaseTransition}Obtaining the phase transition}

In order to obtain a highly-accurate value for the position of the
first order phase transition, we have to find the intersection point
of the localized and the uniform branch in the phase diagrams, cf.
figure~\ref{fig:KKPhasediagrams} and figure~\ref{fig:SYMPhasediagrams}.
For this purpose, we compute a series of localized black hole solutions
with values of the control parameter $\kappa$ that are distributed
on a Lobatto grid around the intersection, i.e. for 
\begin{align}
\kappa_{j} & =\frac{\kappa_{\textrm{end}}+\kappa_{\textrm{start}}}{2}-\frac{\kappa_{\textrm{end}}-\kappa_{\textrm{start}}}{2}\cos\left(\frac{\pi j}{N_{\kappa}}\right)\,,\quad j=0,\ldots,N_{\kappa}\,.\label{eq:kappaLobatto}
\end{align}
An interval $\kappa\in\left[\kappa_{\text{start}},\kappa_{\text{end}}\right]$
in which the corresponding phase transition is located can be easily
identified once the data for producing the phase diagrams is at hand.
We took $\kappa_{\text{start}}=0.98$ and $\kappa_{\text{end}}=1.02$
(recall that we set $L=8$ in our computations). At each of the Lobatto
points we calculate the relevant physical quantities (mass and entropy).
By using standard pseudo-spectral techniques we are able to express
these quantities in the given interval as a truncated Chebyshev series
depending on $\kappa$ with expansion order $N_{\kappa}$. Finally,
we identify the phase transition point by determining the root of
the difference of these functions and the analytically known uniform
branch.

\begin{figure}[ht]
\centering \includegraphics{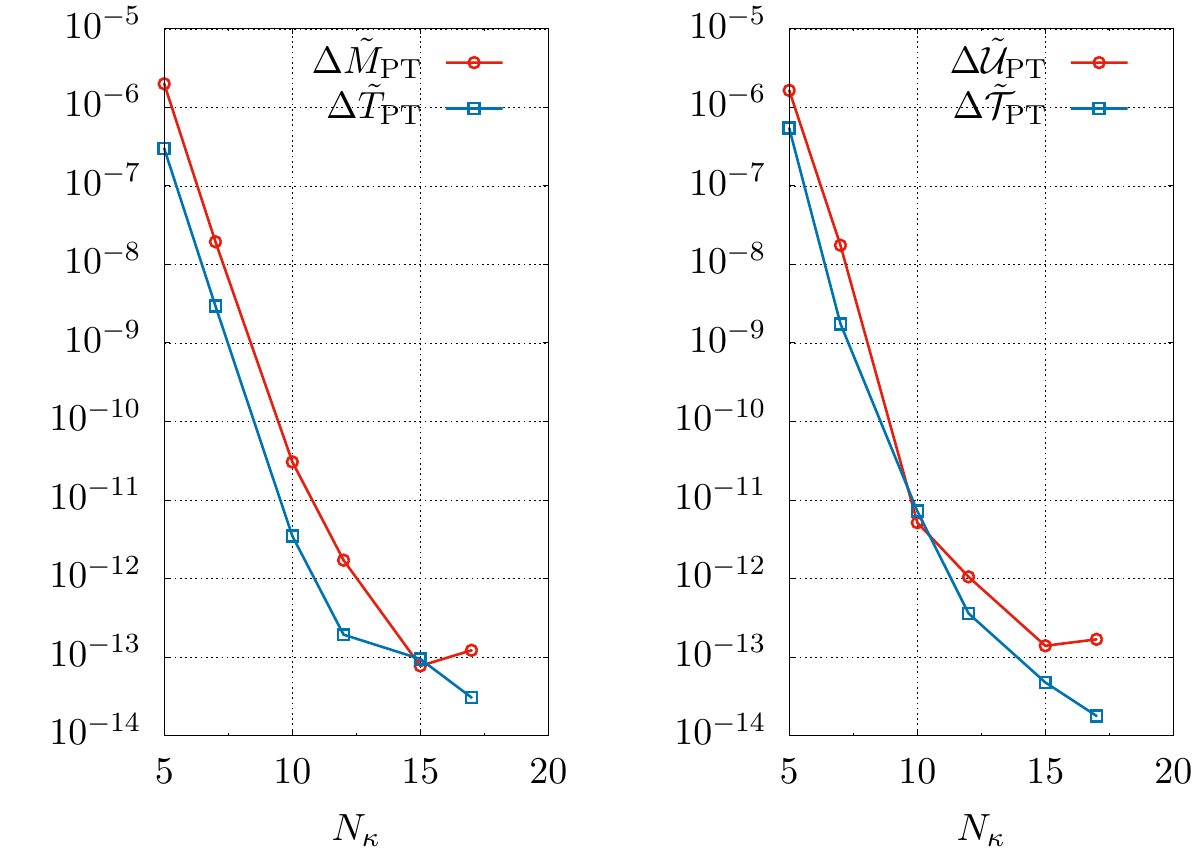} \caption{Convergence of the numerically obtained values of the first order
phase transition between the localized and the uniform phase. We show
the differences to the values of the regarding quantity obtained with
$N_{\kappa}=20$ Lobatto points normalized by this value.}
\label{fig:AccuracyPT} 
\end{figure}

Due to the high accuracy of our localized black hole solutions this
procedure will give highly accurate values for the intersection points
as well, at least if the resolution $N_{\kappa}$ is high enough,
but since we chose a rather small interval $[\kappa_{\text{start}},\kappa_{\text{end}}]$,
comparably small values of $N_{\kappa}$ suffice. Moreover, this approach
provides a natural estimation of accuracy for the phase transition
values, simply by comparing the values obtained for different resolutions,
similarly to the procedure described above. The result of this convergence
analysis is shown in figure~\ref{fig:AccuracyPT}. 

Moreover, this procedure even gives a straightforward way to calculate
derivatives of the thermodynamic quantities in the corresponding interval
$\left[\kappa_{\text{start}},\kappa_{\text{end}}\right]$ by using
standard spectral algorithms, which can be employed for obtaining
the latent heat, cf. equation~(\ref{eq:LatentHeatb}). Besides that,
we are able to check the first law of thermodynamics $\D M = T \, \D S$
in this interval as an additional consistency check of the numerical
results. Indeed, the deviation from this law rapidly drops down as
the resolution is increased and saturates at values of $10^{-12}$.

\section{\label{sec:-SYM-on}Review: $\mathcal{N}=(8,8)$ SYM on $\mathbb{S}^{1}$
and its supergravity description}

Strongly coupled $\mathcal{N}=(8,8)$ SYM in the large-N limit is
conjectured to be equivalent to type II superstring theory, with string
length $l_{s}=\sqrt{\alpha'}$ and string coupling constant $g_{s}$.

As discussed in \cite{Itzhaki:1998dd,Ammon:2015wua}, this duality
may be motivated within string theory by considering $N$ coincident
(non-extremal) D1-branes in type IIB superstring theory. This duality
becomes tractable in the limit $1\ll\lambda\ll N^{4/7}$ which we
assume to hold from now on. In this regime, curvature scales are much
larger than the string scale and the effective string coupling constant
is small. Hence, we can approximate superstring theory by type IIB
supergravity, and the $N$ coincident D1-branes correspond to a particular
(non-extremal) supergravity solution, known as a 1-brane. Finally,
taking the decoupling limit 
\begin{equation}
l_{s}\rightarrow0\quad\textrm{with}\ \ g_{s}l_{s}^{-2}\ \ \text{and}\ \ r/l_{s}^{2}\ \ \text{fixed}\,,
\end{equation}
we arrive at the conjectured duality between $\mathcal{N}=(8,8)$
SYM and type IIB supergravity. Since we also keep $r/l_{s}^{2}$
fixed, where $r$ represents any physical length, we effectively zoom
into the near-horizon part of the supergravity solution.

In addition, to describe $\mathcal{N}=(8,8)$ SYM on a circle $\mathbb{S}^{1}$,
the spatial coordinate of D1-branes has to be compactified on a circle
with circumference $L$. Due to the compactified spatial direction
we have to ensure that the curvature scale has to be small enough
such that stringy excitations winding around the $\mathbb{S}^{1}$
are suppressed. In addition, momentum carrying excitations along the
circle should not excite string oscillations. In the large $N$ limit
and strong coupling limit, this is ensured if $\lambda^{-1/6}\ll\mathcal{T}\ll\sqrt{\lambda}.$
Here $\mathcal{T}$ is the dimensionless temperature associated with
the type IIB supergravity solution and may be identified with the
dimensionless temperature on the field theory side.

Note that the type IIB supergravity solution breaks down for small
enough temperatures. However, for temperatures of order $\lambda^{-1/6}$
or below, we may perform a T-duality along the compactified spatial
direction of the type IIB supergravity solution by using the Buscher
rules \cite{Itzhaki:1998dd,Aharony:2004ig}. In particular, the T-duality
transforms the length of the circle $L$ into $\tilde{L}=4\pi^{2}l_{s}^{2}/L.$
The resulting type IIA supergravity solution is valid for dimensionless
temperatures $\mathcal{T}\ll\lambda^{-1/6}$.

In this paper, we are only interested in the type IIA supergravity
description. This supergravity solution may be viewed as D0-branes
uniformly smeared along the spatial circle $\mathbb{S}^{1}.$ However,
for low enough temperatures, the D0-branes tend to be non-uniformly
smeared along this direction. This instability is reminiscent of the
well-known Gregory-Laflamme instability in asymptotically flat Kaluza-Klein
geometries and was found in \cite{Aharony:2004ig}. In particular,
the onset of the instability occurs for~\cite{Gregory:1993vy,Aharony:2004ig,Dias:2017uyv}
\begin{equation}
\mathcal{\tilde{T}}_{\text{GL}}=\mathcal{T}_{\text{GL}}\,\sqrt{\lambda}\approx2.243\,.
\end{equation}
In \cite{Dias:2017uyv}, the authors construct the associated non-uniform
black string solutions. It is expected that these non-uniform black
string solutions with horizon topology $\mathbb{S}^{1}\times\mathbb{S}^{7}$
will merge into the localized black holes with horizon topology $\mathbb{S}^{8}$.
The latter ones correspond to D0-branes localized on the spatial circle
$\mathbb{S}^{1}.$

\bibliography{main.bbl}

\providecommand{\href}[2]{#2}\begingroup\raggedright\begin{thebibliography}{10}

\bibitem{Gregory:1993vy}
R.~Gregory and R.~Laflamme, ``{Black strings and p-branes are unstable},'' {\em
  Phys. Rev. Lett.} {\bf 70} (1993) 2837--2840,
\href{http://www.arXiv.org/abs/hep-th/9301052}{{\tt hep-th/9301052}}.
%%CITATION = HEP-TH/9301052;%%.

\bibitem{Gregory:1994bj}
R.~Gregory and R.~Laflamme, ``{The Instability of charged black strings and
  p-branes},'' {\em Nucl. Phys.} {\bf B428} (1994) 399--434,
\href{http://www.arXiv.org/abs/hep-th/9404071}{{\tt hep-th/9404071}}.
%%CITATION = HEP-TH/9404071;%%.

\bibitem{Gubser:2001ac}
S.~S. Gubser, ``{On nonuniform black branes},'' {\em Class. Quant. Grav.} {\bf
  19} (2002) 4825--4844,
\href{http://www.arXiv.org/abs/hep-th/0110193}{{\tt hep-th/0110193}}.
%%CITATION = HEP-TH/0110193;%%.

\bibitem{Wiseman:2002zc}
T.~Wiseman, ``{Static axisymmetric vacuum solutions and nonuniform black
  strings},'' {\em Class. Quant. Grav.} {\bf 20} (2003) 1137--1176,
\href{http://www.arXiv.org/abs/hep-th/0209051}{{\tt hep-th/0209051}}.
%%CITATION = HEP-TH/0209051;%%.

\bibitem{Sorkin:2004qq}
E.~Sorkin, ``{A Critical dimension in the black string phase transition},''
  {\em Phys. Rev. Lett.} {\bf 93} (2004) 031601,
\href{http://www.arXiv.org/abs/hep-th/0402216}{{\tt hep-th/0402216}}.
%%CITATION = HEP-TH/0402216;%%.

\bibitem{Kleihaus:2006ee}
B.~Kleihaus, J.~Kunz, and E.~Radu, ``{New nonuniform black string solutions},''
  {\em JHEP} {\bf 06} (2006) 016,
\href{http://www.arXiv.org/abs/hep-th/0603119}{{\tt hep-th/0603119}}.
%%CITATION = HEP-TH/0603119;%%.

\bibitem{Sorkin:2006wp}
E.~Sorkin, ``{Non-uniform black strings in various dimensions},'' {\em Phys.
  Rev.} {\bf D74} (2006) 104027,
\href{http://www.arXiv.org/abs/gr-qc/0608115}{{\tt gr-qc/0608115}}.
%%CITATION = GR-QC/0608115;%%.

\bibitem{Headrick:2009pv}
M.~Headrick, S.~Kitchen, and T.~Wiseman, ``{A New approach to static numerical
  relativity, and its application to Kaluza-Klein black holes},'' {\em Class.
  Quant. Grav.} {\bf 27} (2010) 035002,
\href{http://www.arXiv.org/abs/0905.1822}{{\tt 0905.1822}}.
%%CITATION = ARXIV:0905.1822;%%.

\bibitem{Figueras:2012xj}
P.~Figueras, K.~Murata, and H.~S. Reall, ``{Stable non-uniform black strings
  below the critical dimension},'' {\em JHEP} {\bf 11} (2012) 071,
\href{http://www.arXiv.org/abs/1209.1981}{{\tt 1209.1981}}.
%%CITATION = ARXIV:1209.1981;%%.

\bibitem{Kalisch:2015via}
M.~Kalisch and M.~Ansorg, ``{Highly Deformed Non-uniform Black Strings in Six
  Dimensions},'' in {\em {Proceedings, 14th Marcel Grossmann Meeting on Recent
  Developments in Theoretical and Experimental General Relativity,
  Astrophysics, and Relativistic Field Theories (MG14) (In 4 Volumes): Rome,
  Italy, July 12-18, 2015}}, vol.~2, pp.~1799--1804.
\newblock 2017.
\newblock
\href{http://www.arXiv.org/abs/1509.03083}{{\tt 1509.03083}}.
\newblock
%%CITATION = ARXIV:1509.03083;%%.

\bibitem{Kalisch:2016fkm}
M.~Kalisch and M.~Ansorg, ``{Pseudo-spectral construction of non-uniform black
  string solutions in five and six spacetime dimensions},'' {\em Class. Quant.
  Grav.} {\bf 33} (2016), no.~21, 215005,
\href{http://www.arXiv.org/abs/1607.03099}{{\tt 1607.03099}}.
%%CITATION = ARXIV:1607.03099;%%.

\bibitem{Dias:2017uyv}
O.~J.~C. Dias, J.~E. Santos, and B.~Way, ``{Localised and nonuniform thermal
  states of super-Yang-Mills on a circle},'' {\em JHEP} {\bf 06} (2017) 029,
\href{http://www.arXiv.org/abs/1702.07718}{{\tt 1702.07718}}.
%%CITATION = ARXIV:1702.07718;%%.

\bibitem{Myers:1986rx}
R.~C. Myers, ``{Higher Dimensional Black Holes in Compactified Space-times},''
  {\em Phys. Rev.} {\bf D35} (1987)
455.
%%CITATION = PHRVA,D35,455;%%.

\bibitem{Harmark:2003yz}
T.~Harmark, ``{Small black holes on cylinders},'' {\em Phys. Rev.} {\bf D69}
  (2004) 104015,
\href{http://www.arXiv.org/abs/hep-th/0310259}{{\tt hep-th/0310259}}.
%%CITATION = HEP-TH/0310259;%%.

\bibitem{Gorbonos:2004uc}
D.~Gorbonos and B.~Kol, ``{A Dialogue of multipoles: Matched asymptotic
  expansion for caged black holes},'' {\em JHEP} {\bf 06} (2004) 053,
\href{http://www.arXiv.org/abs/hep-th/0406002}{{\tt hep-th/0406002}}.
%%CITATION = HEP-TH/0406002;%%.

\bibitem{Gorbonos:2005px}
D.~Gorbonos and B.~Kol, ``{Matched asymptotic expansion for caged black holes:
  Regularization of the post-Newtonian order},'' {\em Class. Quant. Grav.} {\bf
  22} (2005) 3935--3960,
\href{http://www.arXiv.org/abs/hep-th/0505009}{{\tt hep-th/0505009}}.
%%CITATION = HEP-TH/0505009;%%.

\bibitem{Wiseman:2002ti}
T.~Wiseman, ``{From black strings to black holes},'' {\em Class. Quant. Grav.}
  {\bf 20} (2003) 1177--1186,
\href{http://www.arXiv.org/abs/hep-th/0211028}{{\tt hep-th/0211028}}.
%%CITATION = HEP-TH/0211028;%%.

\bibitem{Sorkin:2003ka}
E.~Sorkin, B.~Kol, and T.~Piran, ``{Caged black holes: Black holes in
  compactified space-times. 2. 5-d numerical implementation},'' {\em Phys.
  Rev.} {\bf D69} (2004) 064032,
\href{http://www.arXiv.org/abs/hep-th/0310096}{{\tt hep-th/0310096}}.
%%CITATION = HEP-TH/0310096;%%.

\bibitem{Kudoh:2003ki}
H.~Kudoh and T.~Wiseman, ``{Properties of Kaluza-Klein black holes},'' {\em
  Prog. Theor. Phys.} {\bf 111} (2004) 475--507,
\href{http://www.arXiv.org/abs/hep-th/0310104}{{\tt hep-th/0310104}}.
%%CITATION = HEP-TH/0310104;%%.

\bibitem{Kudoh:2004hs}
H.~Kudoh and T.~Wiseman, ``{Connecting black holes and black strings},'' {\em
  Phys. Rev. Lett.} {\bf 94} (2005) 161102,
\href{http://www.arXiv.org/abs/hep-th/0409111}{{\tt hep-th/0409111}}.
%%CITATION = HEP-TH/0409111;%%.

\bibitem{Kalisch:2017bin}
M.~Kalisch, S.~Möckel, and M.~Ammon, ``{Critical behavior of the black
  hole/black string transition},'' {\em JHEP} {\bf 08} (2017) 049,
\href{http://www.arXiv.org/abs/1706.02323}{{\tt 1706.02323}}.
%%CITATION = ARXIV:1706.02323;%%.

\bibitem{Kol:2004ww}
B.~Kol, ``{The Phase transition between caged black holes and black strings: A
  Review},'' {\em Phys. Rept.} {\bf 422} (2006) 119--165,
\href{http://www.arXiv.org/abs/hep-th/0411240}{{\tt hep-th/0411240}}.
%%CITATION = HEP-TH/0411240;%%.

\bibitem{Harmark:2005pp}
T.~Harmark and N.~A. Obers, ``{Phases of Kaluza-Klein black holes: A Brief
  review},''
\href{http://www.arXiv.org/abs/hep-th/0503020}{{\tt hep-th/0503020}}.
%%CITATION = HEP-TH/0503020;%%.

\bibitem{Horowitz:2011cq}
G.~T. Horowitz and T.~Wiseman, ``{General black holes in Kaluza-Klein
  theory},'' in {\em {Black holes in higher dimensions}}, G.~T. Horowitz, ed.
\newblock Cambridge University Press, Cambridge, UK, 2012.
\newblock
\href{http://www.arXiv.org/abs/1107.5563}{{\tt 1107.5563}}.
\newblock
%%CITATION = ARXIV:1107.5563;%%.

\bibitem{Kalisch:2018efd}
M.~Kalisch, {\em {Numerical construction and critical behavior of Kaluza-Klein
  black holes}}.
\newblock PhD thesis, Jena U., 2018.
\newblock
\href{http://www.arXiv.org/abs/1802.06596}{{\tt 1802.06596}}.
\newblock
%%CITATION = ARXIV:1802.06596;%%.

\bibitem{Kol:2002xz}
B.~Kol, ``{Topology change in general relativity, and the black hole black
  string transition},'' {\em JHEP} {\bf 10} (2005) 049,
\href{http://www.arXiv.org/abs/hep-th/0206220}{{\tt hep-th/0206220}}.
%%CITATION = HEP-TH/0206220;%%.

\bibitem{Kol:2005vy}
B.~Kol, ``{Choptuik scaling and the merger transition},'' {\em JHEP} {\bf 10}
  (2006) 017,
\href{http://www.arXiv.org/abs/hep-th/0502033}{{\tt hep-th/0502033}}.
%%CITATION = HEP-TH/0502033;%%.

\bibitem{Itzhaki:1998dd}
N.~Itzhaki, J.~M. Maldacena, J.~Sonnenschein, and S.~Yankielowicz,
  ``{Supergravity and the large N limit of theories with sixteen
  supercharges},'' {\em Phys. Rev.} {\bf D58} (1998) 046004,
\href{http://www.arXiv.org/abs/hep-th/9802042}{{\tt hep-th/9802042}}.
%%CITATION = HEP-TH/9802042;%%.

\bibitem{Ammon:2015wua}
M.~Ammon and J.~Erdmenger, {\em {Gauge/gravity duality}}.
\newblock Cambridge University Press,
2015.
\newblock
%%CITATION = INSPIRE-1376202;%%.

\bibitem{Aharony:2004ig}
O.~Aharony, J.~Marsano, S.~Minwalla, and T.~Wiseman, ``{Black hole-black string
  phase transitions in thermal 1+1 dimensional supersymmetric Yang-Mills theory
  on a circle},'' {\em Class. Quant. Grav.} {\bf 21} (2004) 5169--5192,
\href{http://www.arXiv.org/abs/hep-th/0406210}{{\tt hep-th/0406210}}.
%%CITATION = HEP-TH/0406210;%%.

\bibitem{Harmark:2004ws}
T.~Harmark and N.~A. Obers, ``{New phases of near-extremal branes on a
  circle},'' {\em JHEP} {\bf 09} (2004) 022,
\href{http://www.arXiv.org/abs/hep-th/0407094}{{\tt hep-th/0407094}}.
%%CITATION = HEP-TH/0407094;%%.

\bibitem{Catterall:2010fx}
S.~Catterall, A.~Joseph, and T.~Wiseman, ``{Thermal phases of D1-branes on a
  circle from lattice super Yang-Mills},'' {\em JHEP} {\bf 12} (2010) 022,
\href{http://www.arXiv.org/abs/1008.4964}{{\tt 1008.4964}}.
%%CITATION = ARXIV:1008.4964;%%.

\bibitem{Catterall:2017lub}
S.~Catterall, R.~G. Jha, D.~Schaich, and T.~Wiseman, ``{Testing holography
  using lattice super-Yang-Mills theory on a 2-torus},'' {\em Phys. Rev.} {\bf
  D97} (2018), no.~8, 086020,
\href{http://www.arXiv.org/abs/1709.07025}{{\tt 1709.07025}}.
%%CITATION = ARXIV:1709.07025;%%.

\bibitem{Jha:2017zad}
R.~G. Jha, S.~Catterall, D.~Schaich, and T.~Wiseman, ``{Testing the holographic
  principle using lattice simulations},'' {\em EPJ Web Conf.} {\bf 175} (2018)
  08004,
\href{http://www.arXiv.org/abs/1710.06398}{{\tt 1710.06398}}.
%%CITATION = ARXIV:1710.06398;%%.

\bibitem{Pau}
B.~Cardona and P.~Figueras, ``{Critical Kaluza-Klein black holes and black
  strings in D=10},'' \href{http://www.arXiv.org/abs/to appear}{{\tt to
  appear}}.

\bibitem{Emparan:2018bmi}
R.~Emparan, R.~Luna, M.~Martinez, R.~Suzuki, and K.~Tanabe, ``{Phases and
  Stability of Non-Uniform Black Strings},''
\href{http://www.arXiv.org/abs/1802.08191}{{\tt 1802.08191}}.
%%CITATION = ARXIV:1802.08191;%%.

\bibitem{Wiseman:2011by}
T.~Wiseman, ``{Numerical construction of static and stationary black holes},''
  in {\em {Black holes in higher dimensions}}, G.~T. Horowitz, ed.
\newblock Cambridge University Press, Cambridge, UK, 2012.
\newblock
\href{http://www.arXiv.org/abs/1107.5513}{{\tt 1107.5513}}.
\newblock
%%CITATION = ARXIV:1107.5513;%%.

\bibitem{Dias:2015nua}
O.~J.~C. Dias, J.~E. Santos, and B.~Way, ``{Numerical Methods for Finding
  Stationary Gravitational Solutions},'' {\em Class. Quant. Grav.} {\bf 33}
  (2016), no.~13, 133001,
\href{http://www.arXiv.org/abs/1510.02804}{{\tt 1510.02804}}.
%%CITATION = ARXIV:1510.02804;%%.

\bibitem{Figueras:2011va}
P.~Figueras, J.~Lucietti, and T.~Wiseman, ``{Ricci solitons, Ricci flow, and
  strongly coupled CFT in the Schwarzschild Unruh or Boulware vacua},'' {\em
  Class. Quant. Grav.} {\bf 28} (2011) 215018,
\href{http://www.arXiv.org/abs/1104.4489}{{\tt 1104.4489}}.
%%CITATION = ARXIV:1104.4489;%%.

\bibitem{Kol:2003if}
B.~Kol, E.~Sorkin, and T.~Piran, ``{Caged black holes: Black holes in
  compactified space-times. 1. Theory},'' {\em Phys. Rev.} {\bf D69} (2004)
  064031,
\href{http://www.arXiv.org/abs/hep-th/0309190}{{\tt hep-th/0309190}}.
%%CITATION = HEP-TH/0309190;%%.

\bibitem{Harmark:2003dg}
T.~Harmark and N.~A. Obers, ``{New phase diagram for black holes and strings on
  cylinders},'' {\em Class. Quant. Grav.} {\bf 21} (2004) 1709,
\href{http://www.arXiv.org/abs/hep-th/0309116}{{\tt hep-th/0309116}}.
%%CITATION = HEP-TH/0309116;%%.

\end{thebibliography}\endgroup

\end{document}